# Mining Treatment-Outcome Constructs from Sequential Software Engineering Data

Maleknaz Nayebi, *Member, IEEE*, Guenther Ruhe, *Senior Member, IEEE* and Thomas Zimmermann
*Senior Member, IEEE*

**Abstract**—Many investigations in empirical software engineering look at sequences of data resulting from development or management processes. In this paper, we propose an analytical approach called the *Gandhi-Washington Method (GWM)* to investigate the impact of recurring events in software projects. GWM takes an encoding of events and activities provided by a software analyst as input. It uses regular expressions to automatically condense and summarize information and infer treatments. Relating the treatments to the outcome through statistical tests, treatment-outcome constructs are automatically mined from the data. The output of GWM is a set of treatment-outcome constructs. Each treatment in the set of mined constructs is significantly different from the other treatments considering the impact on the outcome and/or is structurally different from other treatments considering the sequence of events. We describe GWM and classes of problems to which GWM can be applied. We demonstrate the applicability of this method for empirical studies on sequences of file editing, code ownership, and release cycle time.

---

## 1 INTRODUCTION

SOFTWARE development deals with a large amount of data created from sequential activities, events, and decisions. Many investigations in empirical software engineering look for sequences of events as the independent variables and analyze the related outcomes for all these variations. With the increasing amount of results gained from empirical investigations in software engineering, there is a substantial need to structure and synthesize this knowledge. We propose a method for empirical studies where the "cause constructs" [50] are "sequences of activities, events or decisions".

Developers change the source code, perform code reviews, commit code, run builds and test cases, and they iteratively release software product versions. Requirements engineers have access to sequential data from comprehensive user feedback, usage data from social media, forums, and review systems [23]. Customers use the software, navigate through features, and submit bug reports or feature requests. Many of the questions that software engineers have are related to such sequence of events, activities, and decisions. As a result, questions like "Should developers commit code before reviewing the changes?" or "should they review changes before committing the code?" are asked and investigated empirically [40], [41]. Therein, commit-then-review and review-then-commit represent sequences of review and commit activities.

- M. Nayebi is with Ecole Polytechnique of Montreal, Canada
  E-mail: mnayebi@polymtl.ca
- G. Ruhe is with the Software Engineering Decision Support Laboratory, University of Calgary, Canada
  E-mail: ruhe@ucalgary.ca
- T. Zimmermann is with Microsoft research. E-mail: tzimmer@microsoft.com

The principle and process of experimentation as adapted from [50] is shown in Figure 1. For performing an empirical study in software engineering, we first form a hypothesis around cause and effect. Then, we test that hypothesis against our observations by conducting an experiment (see Figure 1). In this paper, we propose a method exclusively for testing hypotheses that considers sequences of activities as treatments. *Treatment* is often referred to an intervention, which is a method or an independent variable that causes some measurable factor to change [21]. The result of our method is a set of TrOC's that describe a significant impact of treatments on the outcome Proposing a methodology for mining and analyzing the impact of event sequences is the objective of this research.

We introduce an analytical approach called the *Gandhi-Washington Method*[1] or shortly GWM. The basic steps of GWM can be summarized into three phases:

1) **Encoding.** A software analyst (or engineer) encodes a sequence of events (we later refer to them as items) into a sequence of characters taken from an alphabet of her choice.
2) **Abstraction.** An automated step that summarizes the encoded sequences by using regular expressions.
3) **Synthesis.** An automated step that applies statistical tests on the regular expressions. This step merges the categories that have an insignificant relationship with the specified outcome.

The output of GWM is a set of treatment-outcome constructs [50] or *TrOC* in short. TrOC's are regular expressions that show the condensed and essential sequence of events (treatments). These regular expressions are associated with the values of a specified dependent variable (outcome). GWM is a semi-automated approach. Once an analyst encodes the problem and selects the desirable performance measures, a series of statistical tests automatically retrieve treatment-outcome constructs (TrOCs).

---

[1]. The name is inspired by the homological diversity and discipline that Gandhi and Washington both stood for. This is a metaphor for the alternative structuring and occurrence of events when developing a software product.

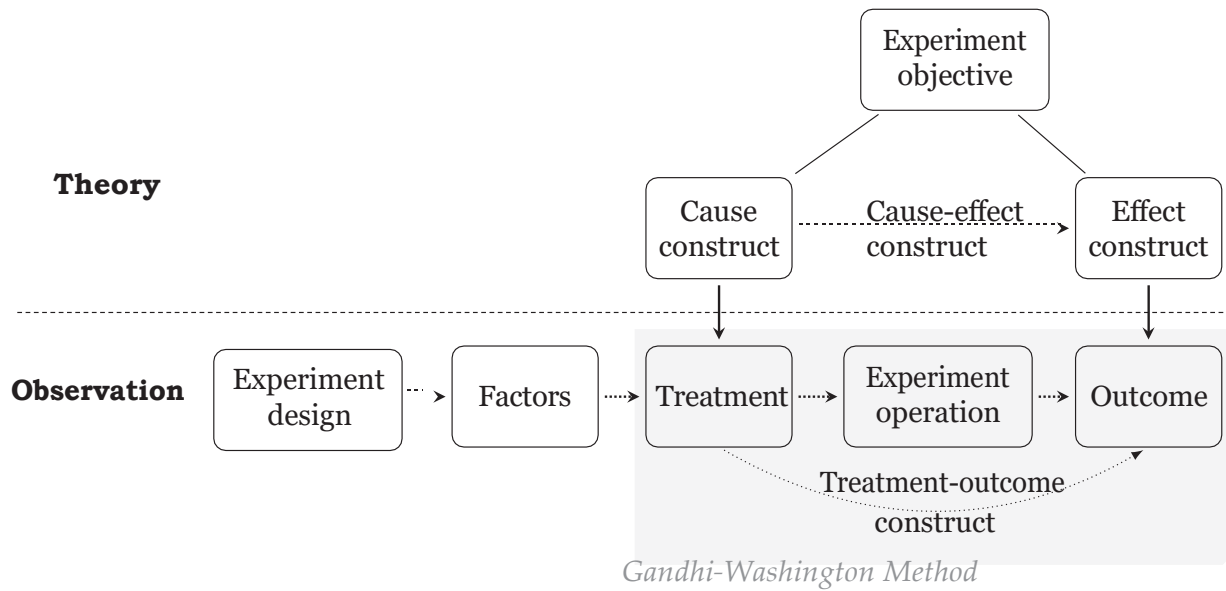

Fig. 1. The principle and process of experimentation as adopted from Wohlin et al. [50]. GWM aims at mining the relation between treatment and outcome and is located in the domain of observation. Cause and effect constructs forms the hypothesis in the "theory space. For our study, the relation between the cause construct and the treatment as well as the relation between the effect construct and the outcome is one-directional. The theory provides the hypothesis for modeling the problem to be examined in the observation space. The treatment-outcome construct does not show causality but the relation between these two in a specific case study context.

For example, an analyst hypothesized that "the sequence of commit and review activities have an impact on the number of bugs in the code". She designed an experiment and observed the relation between two treatments "commit-then-review" and "review-then-commit" and defined outcome as the "number of bugs". She made the treatment-outcome construct that "review-then-commit is related to the significantly less number of bugs comparing to the commit-then-review sequence". However, by having GWM, the analyst only needs to encode the activities commit and review (for example by encoding "commit" by "C" and "review" by "R") and provides a dataset of the experiment. GWM then automatically mines treatments and their relation to the outcome.

The contribution of this paper is the Gandhi-Washington Method. GWM is a general approach to analyze the relationship between sequential events in software processes or products and an outcome. GWM can be applied to a large class of software engineering decision problems related to sequences of events. In empirical software engineering, researchers often evaluate alternative sequences and measure the related outcomes. GWM facilitates this process by (i) providing a unified method to mine treatments from data (ii) synthesizing treatments based on the impact on a specific outcome (iii) making the process easily replicable. In Figure 1, the role of GWM in the process of running empirical studies is illustrated. To apply GWM for a new analysis problem, a developer or analyst first has to model the problem by adjusting the encoding. The remaining steps are tool supported (see Section 5).

In this paper we first introduce and motivate GWM. To illustrate the applicability of GWM, we provide two example scenarios to motivate our research. In Section 3, we describe the details of GWM, followed by Section 4 in which we discuss three applications of GWM. Prototype tool support is the content of Section 5. Implementation aspects such as multiple encoding and computational complexity are elaborated in Section 6. Limitations of the applicability of GWM are discussed in Section 7. In Section 8, we present related work and compare existing sequential data analysis methods with GWM. Finally, we conclude the paper in Section 9.

## 2 MOTIVATION

We present two motivating scenarios to explain the problem under consideration. The first scenario relates to the relationship between code healthiness and code review practices. The second scenario looks into the relation between the number of conflicts and sequence of code verification periods.

**Scenario #1:** Project manager Alice is looking into ways to enhance the healthiness of the code developed by her team. The team doesn't have a unified way of committing, testing and reviewing the code. Some of the sub-project teams follow a commit-test-review process, while others apply review-test-commit. The rest does not follow either of these strategies to review their code. Alice wonders if the selection of one of these strategies affects the number of code bugs. To answer this question, she selects GWM and encodes each method for testing and reviewing a file by the sequence of testing (T), committing (C), and reviewing

TABLE 1
Applying Gandhi-Washington Method on Scenario #2.

| Step 1: Encoding (Analyst) | |
|---|---|
| *What to encode?* | The length of verification periods. |
| *How to encode?* | Discretize the length into short (S), medium (M) and long (L) periods. |
| *What to select as performance measure?* | Number of conflicts. |
| *Sample encoded periods (and their number of conflicts)* | $SSSSSS$ (23), $LLLLLLLLLL$ (47), $MMSMMS$ (73), $SSSSLLLLLL$ (52), $MSLMSS$ (66). |
| **Step 2: Abstraction (Automated step in GWM)** | |
| *What to abstract?* | Abstract the encoded verification periods. |
| *How to abstract?* | Use regular expressions to classify encoded strings. |
| *Sample abstracted strings* | $S*$, $L*$, $(M*S)*$, $S*L*$, $(MS*L*)*$. |
| **Step 3: Synthesis (Automated step in GWM)** | |
| *What to synthesize?* | Classes of regular expressions. |
| *How to synthesize?* | Apply the MannWhitney test to compare regular expressions if they are related to significant difference in the number of conflicts. |
| *Sample output of GWM* | Set of TrOCs (such as $L*$, which represents consecutive long verification periods) indicating unique sequences of events with differing mean number of conflicts. |

(R) activities on it. She also selects *number of bugs* as the outcome.

**Cause-effect construct:** sequence of activities for testing, committing, and reviewing the code has an impact on the number of code bugs.
**Outcome:** number of bugs.
**Treatments:** TCR, TRC, CRT, CTR, RTC, RCT.

Applying GWM, she observes that pieces of code which are alternating CRT sequence for commit, test, and review have a significantly fewer number of bugs compared to the other approaches.

**Scenario #2:** Product manager Bob wants to define a strategy for the periodic verification of the code's trunk branch (as studied by Murphy et al. [30]). In each iteration, a number of commits are integrated to the trunk branch. After a period of time, the product is built and verified. The product manager, trying to reduce the risk of large merge conflicts, defined a strategy of verification being fixed to two days time intervals. However, he also considered to delay the verification by three more days due to difficulties in running the example project. Bob is concerned about the impact of this unscheduled variation. He hypothesized that a fixed and very short verification period decreases the number of conflicts in the code. To get an answer, Bob collects the data from all the former projects and performs an analysis. For this scenario, we compare two analysis approaches: Traditional method which does not use GWM versus using GWM.

*Scenario #2 - Traditional approach:* Bob hypothesized that a sequence of fixed and very short verification periods decreases the number of conflicts (outcome). He defined two treatments. First, a sequence of fixed and very short verification periods and second, all other variations of the verification periods. He divides the former projects into two subsets: The first group contains the projects that had fixed short verification periods and the second group all the other projects. He then compares these two groups by running the Mann–Whitney test on the number of conflicts per project.

*Scenario #2 - Using GWM:* Bob encodes the verification periods into short (S), medium (M), and long (L). He applies GWM as shown in Table 1 for the purpose to validate the cause-effect construct. The results of GWM show that $L*$ (consecutive long verification periods), $(SL)*$ (consecutive pairs of short and long release cycles), and all the rest of processes $(S*M*L*)*$ have a significantly different (and higher) impact on the number of conflicts. The use of GWM helps to define and run a variety of encodings as the number of items would not have a visible impact on the effort to run the experiment. So, in this example, Bob decides to define three encodings rather than two.

**Cause-effect construct:** length and variation of verification periods has impact on the number of conflicts.
**Outcome:** number of conflicts.
**Treatments:** all the variations and combinations of S, L, M.

Having GWM at hand, Bob has more flexibility to test the hypothesis as he can encode verification periods according to the specific context. So he can introduce a "medium" verification period as well without complication in the design of the experiment.

In both scenarios, the analysis is concerned with the sequence of events that are happening in the project and how these sequences relate to the specified outcomes. The managers can make their own model from the system and analyze the problems [36] by adopting some proposed models for code review practices and inter-team coordination [5] in Scenario #1 and branching strategies for scenario #2 [44]. However, having a semi-automatic method such as GWM makes obtaining such experiences less time consuming and more accurate to perform.

## 3 GANDHI-WASHINGTON METHOD

In this section we introduce the Gandhi-Washington Method (GWM). We define *items* as the basic information for our mining process. *Items* are software related activities and events that are stored as transactions in the software repositories. *Itemsets* are groups of items that occur together in sequence and are grouped by a time-stamp [24]. Within this paper, as long as we are discussing itemsets, we are referring to *ordered* itemsets. In the context of GWM, treatment-outcome constructs (TrOC) are recurring sequences of events (treatments) that are statistically related to software product measures (outcomes).

GWM uses the notion of regular expressions and their grammatical relation within a formal language to summarize the structure of itemsets. Next, observations are synthesized by using statistical tests to extract structure of sequences that affect a specific dependent variable in the software or process (outcome). The selected dependent variable is called the *outcome*. The rest of this section discusses the three phases of Gandhi-Washington Method in detail.

### 3.1 Encoding

A formal language $L$ over an alphabet $\Sigma$ is a set of all strings permitted by the rules of formation and is a subset of $\Sigma^*$, where $\Sigma^*$ is the set of all possible combinations (considering sequence) of the letters over the alphabet. In this context, the *cardinality of* $\Sigma$ is the number of items encoded in an itemset. Ordered itemsets can also be presented as a string of items. A regular expression provides a structure to express a class of strings. The length of a string is the number of characters within that string (for example, the length of |$AABA$| is 4).

Each event or event type can be assigned to a letter. A big portion of sequential data in software engineering is on a nominal or ordinal scale. Encoding assigns one character to all the items (events) of similar type. For example, in the Scenario #2 (verification patterns of repository trunk branches), |$S$| is used to encode all the verification periods which had a length between 1 to 4 days. Software development is a series of events (items). For synthesizing results, attributes on other scales can be mapped into an ordinal scale. An analyst should identify the types of events in the processes and projects (problem space) and assigns a letter from $\Sigma$ to each event. For example, in Scenario #2 (verification patterns of repository trunk branches), |$S$| is used to encode all the verification periods which had a length between 1 to 4 days. Encoding, as a form of problem modeling, requires a deep understanding of the problem space. In other words, encoding is the art of preparing a model for applying abstraction and synthesis. Here, an encoded itemset is formed by assigning a symbol to each item.

**Categorical items:** Categorical (nominal) data include a fixed number of possible values. For example, if the people involved in the commit-review process is of interest (like Alice, Bob, and Carol), we assign one letter to each, i.e. |$A$| for Alice, |$B$| for Bob and |$C$| for Carol. Then the Alice, Carol, and Bob process are encoded in the form of |$ACB$|. If their organizational position in the process is of interest and having Alice as developer and Bob and Carol as code reviewers, we assign |$D$| to Alice and |$R$| to Bob and Carol so the Alice, Carol, and Bob process are encoded as |$DRR$|.

**Non-categorical data:** While continuous data can range between negative and positive infinity, only some important distinctions might be of interest [26]. Discretization creates different groups (bins) of data. For encoding ordinal data, different discretization or clustering algorithms can be used [26]. Also, experts opinions and their perception of groups in the data could be used for discretization. In the Alice, Carol, and Bob example, we are interested in the number of bugs each of them reported as their performance measure (outcome). We categorize the number of reported issues between 0 and 4 as low ($L$), 5 to 10 as medium ($M$), and more than 10 as high ($H$). If Alice, Bob, and Carol reported 2, 11, and 14 bugs respectively (itemset looks like 2, 11, 14), then the encoded itemset is |$LHH$|.

### 3.2 Abstraction

The abstraction receives encoded itemsets as the input and groups these itemsets into different classes with regards to the sequential commonality between them. Each of these classes is represented by a regular expression. In this step, we move from encoded strings to the set of strings represented by a regular expression. So instead of focusing on the actual strings such as ABB or ABBBBBB we focus on the sequence of items' occurrence in the form of AB...B.

Regular expressions provide a compact view of an itemset and bring focus to the occurrence of sequences. In a regular expression, the use of the Kleene star (*) shows zero or more occurrences of an item type. Following this, both encoded strings $ABBBBBB$ and $ABB$ are categorized by the regular expression $AB^*$. From transforming all itemsets into a regular expression, an enumerated set of regular expressions is formed over a formal language $\Sigma$ (see Figure 2). Depth-First Search (DFS) [11] finds the proper regular expression for each itemset within the hierarchy.

Regular expressions have structural differences (for example, $AB^* \neq B^*A$). Some regular expressions are positioned as the parent of the others. A regular expression is the parent of another one if it can produce the strings of the child regular expression. The parent-child relation is expressed by an edge in the hierarchy. Figure 2 shows the hierarchy of regular expressions for the alphabet $\Sigma = \{A, B\}$. When comparing two regular expressions within a hierarchy, the greater the distance of a node from the root node, the more specific the regular expression. Consequently, each node is more specific than its parent and thus can produce less variety of sequences.

The hierarchy of regular expressions could be extended by using production rules. Production rules replace $A^*$ with $A...AA^*$ and $B^*$ with $B...BB^*$ so an infinite number of strings and children can be produced for each regular expression. In this way, a more detailed hierarchy could be defined over $\Sigma = \{A, B\}$. Substitution of symbols can be recursively performed to generate diverse regular expressions. The more details required for the analysis, the more such production rules could be applied to expand the hierarchies over each alphabet. Among all possible regular expressions and hierarchies, we only considered those expressions created from (i) the letters of $\Sigma$ with no repetition allowed, (ii) applying Kleene star with the number of Kleene stars between zero and $|\Sigma|+1$, (iii) using brackets to define the application of Kleene star.

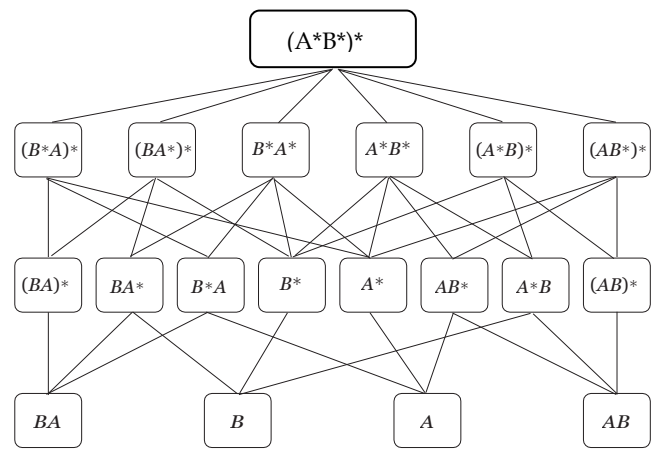

Fig. 2. A hierarchy of regular expressions over $\Sigma = \{A, B\}$.

As the result, we ended up in a finite set of regular expressions. Each itemset is a string of letters from an alphabet. By having a hierarchy of regular expressions over different sets of an alphabet, each encoded itemset is abstracted into the most specific regular expression. To categorize the encoded itemset in a hierarchy of regular expressions, a search is started from the leaf node with the highest distance from the root. Each encoded itemset is compared with the hierarchy nodes until a match is found. One encoded itemset might match with more than one regular expression when two regular expressions have parent-child relation. GWM always classifies each encoded itemset in the most specific regular expression possible. In this way, itemset $\{A, A, A, A, B\}$ is categorized in $A*B$ and not in $A*B*$ even though both regular expressions match the itemset.

As the result of this step, each itemset is categorized in one regular expression and we moved from the actually encoded itemsets into the regular expressions that abstracted sequence of items. Looking into the sample regular expression hierarchy in Figure 2, some nodes of the hierarchy might remain empty after abstraction. That means that no such sequence occurred in the existing data. Non-empty regular expressions are the treatments for our empirical investigations. Treatments are the input of the GWM synthesis process.

### 3.3 Synthesis

The synthesis studies items that are categorized in the hierarchy of regular expressions. This includes the application of a series of statistical tests on a selected dependent variable (outcome). Synthesis step analyzes the commonality of recurring sequences and merges hierarchy nodes based on the results. Merging two nodes in this context means to systematically transfer strings that are classified with a specific regular expression into a more general regular expression in the hierarchy and changing the level of abstraction of the transferred itemsets. Merging is applied based on the impact of recurring sequences on a specified outcome.

The outcome is the dependent variable that describes the effect of the recurring sequences (treatments). The outcome can represent various aspects of software development ranging from code related issues (such as the number of bugs, the number of commits, etc.) to market-related concerns (such as market share of a product, number of downloads, etc.). The Algorithm start by the nodes with the highest distance "i" from the root (Line 4 of Algorithm 1).

**Synthesis Algorithm:** Algorithm 1 shows the synthesis process. The order of the process is following depth first search (DFS) [11].

---

**Algorithm 1:** Synthesis

**Input:** Hierarchy of regular expressions
1  MW[x][y] = Mann–Whitney test between node x and y;
2  Var i = Maximum distance in a tree from root;
3  **Func** Synthesize(Regular expression hierarchy)
4  **while** $i \geq 0$ **do**
5     **while** *Not all the nodes are decided or merged* **do**
6        **if** *a node does not have any siblings* **then**
7           Select it as NODE ;
8        **else if** *a node has one parent only* **then**
9           Select it as NODE ;
10       **else**
11          Select the node with most number of not rejected $H_0$ as NODE;
12          **if** *multiple nodes could be selected* **then**
13             Perform depth first search and select the least significant node as NODE
14          **end**
15    **end**
16    Check and adjust p-value correction;
17    MW[NODE][$NODE_{Parents}$];
18    MW[NODE][$NODE_{Siblings}$];
19    **if** *NODE rejects all $H_0$* **then**
20       **for** *all the NODE siblings* **do**
21          MW[$Node_{sibling}$][$Node_{sibling_{Parents}}$];
22          MW[$Node_{sibling}$][$Node_{sibling_{Siblings}}$];
23          Compare all the siblings and select the least significant node as NODE;
24       **end**
25       Merge NODE with parent returning the highest p-value and smallest effect size ;
26       **if** *NODE has decided children* **then**
27          Transmit the children into NODE parent;
28          Update Mann–Whitney test results;
29       **end**
30    **else if** *All the children and siblings are decided* **then**
31       Mark the NODE as decided;
32    **else**
33       Continue;
34    **end**
35   **end**
36   i = i -1 ;
37 **end**

---

However, it includes exceptions for the nodes without a child, a sibling, or multiple parents. At each step of this process, among the nodes with the greatest distance from the root, the node with the highest priority is selected. The Mann-Whitney test is performed two by two and for all the nodes[2]. If not stated otherwise, we applied the statistical tests with a significance level of 0.05. For the entire algorithm, nodes with no siblings and only one parent have higher priority than others. When all such nodes are analyzed, the nodes with the greatest number of not rejected null hypotheses are selected. When we test the relationship of a node and its siblings based on outcome and decide not to merge that node, we call this node a significant node. If this significant node is not merged with any other nodes during the recursive synthesis, the final regular expression related to the node shows the TroC. For the rest of this section, we refer to lines of Algorithm 1 and describe it in depth.

---

2. In Section 6.2, we elaborate on the possibility to use alternative tests.

**Order in the synthesis:** *(Algorithm 1 - Line 6-19)* The synthesis process is applied on all the hierarchy of nodes until all of them have been statistically compared with their parents and siblings. To synthesize structures and extract treatments with a significant impact on the outcome, the node with the greatest distance from the root that satisfies the following conditions is selected:

1) The node has not been decided in the synthesis. This means that no decision was made for merging or not merging the node with its parent.

2) The node that does not have any child or any undecided child.

3) Between nodes with the highest distance, the priority is given to the node that does not have siblings and then to the node that has only one parent. This says:
   (i) If a node has no siblings, the first if statement (lines 6 to 8) defines the NODE.
   (ii) If a node has one parent only the second if statement (lines 8 and 10) defines the NODE.
   (iii) In the other cases the else statement (Lines 10 to 15) defines the NODE. In cases that there are multiple node candidates, we select the NODE using DFS [11] order.

**Synthesis – Base case:** *(Algorithm 1, Lines 16-18, and 25)* Considering a node with one parent and one sibling we discuss a simple synthesis process. First, the synthesis process compares the node with its parent. The synthesis merges nodes with the parent only in the case where no significant difference between the node and its sibling or its parent is detected in terms of the selected outcome. Assuming that outcome is ordinal, the non-parametric analysis of variance, Mann–Whitney is used in synthesis. The Mann–Whitney test provides a comparison between all the itemsets that GWM categorized in two nodes. The output of the synthesis is a set of sequences that have a significant relationship with the outcome. This is called *TrOC (treatment-outcome construct)*. We are interested in the comparisons between NODE and its parents as well as between NODE and its siblings (Line 17 and 18 of Algorithm 1).

However, for deciding on a node that has siblings, Mann-Whitney test compares the node with all of its siblings. If the null hypothesis isn't rejected and the results show less significance (higher p-value) in terms of outcome compared to its siblings, the node is merged with its parent. In the case that both show insignificance, the node with a smaller effect size is merged first (Line 24 of Algorithm 1).

In what follows, the simplest synthesis on a node which has one parent and one sibling only is explained. Within this process, "applying Mann-Whitney test" refers to comparing the outcomes of all the itemsets categorized in a regular expression. Considering the "NODE" and using the p-value of the statistical tests, we show the application of synthesis:

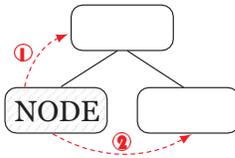

**Synthesis: Base case**

Apply Mann-Whitney test to parent and siblings.

Continuous and ordinal data ⇒
**If** $H_0$ isn't rejected in at least one of the tests (①or②), merge the NODE with its parent.

⇒ **Else**, keep the NODE unmerged.

**Synthesis – Comparison of a node with its parents:** *(Algorithm 1, Lines 17, 21)* Nodes with more than one parent in the regular expression hierarchy need an extension of the synthesis base case. The extended synthesis uses the results of the statistical test to decide on merging or not merging the node. The process for a node with two parents (considering "NODE") is explained below.

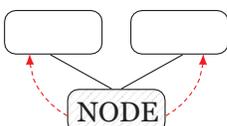

**Synthesis of a node with two parents**

Apply Mann–Whitney tests between NODE and its parents.

⇒ **If** both tests do not reject $H_0$, NODE is merged with the parent returning the higher p-value from the Mann–Whitney test.

⇒ **Else**, the node is not merged.

**Synthesis – Child transmission in merging:** *(Algorithm 1, Lines 26-29)* In the synthesis, if we merge a node which itself has a significant child node, the significant child node would be transmitted to the new combined parent and its significance would be reevaluated considering its new position in the tree. Considering "NODE" in this situation we describe synthesis below:

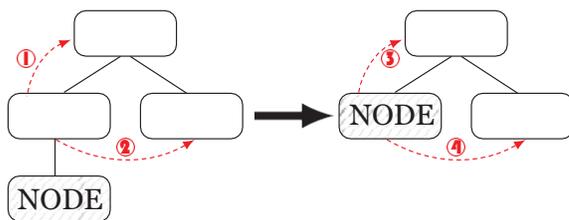

**Transmit Significant Nodes**

Apply Mann–Whitney test between NODE and its parent and its sibling.
1- Merge NODE parent with its parent.
2- Transfer NODE to the new parent.
3- Apply Mann–Whitney test to new parent.
4- If $H_0$ isn't rejected, apply Mann–Whitney to the sibling.

⇒ **If** the $H_0$ isn't rejected merge the NODE with parent.

It is worth to note that, in the synthesis process if a node is empty the Mann-Whitney test would not be able to reject the null hypothesis and the algorithm would continue by merging as described.

**Output:** GWM determines a set of TrOC, $T_1$ to $T_L$, which meet the following conditions:

1) Each itemset is categorized in exactly one TrOC.
2) $T_i \neq T_j \quad \forall i, j$.
3) Mann–Whitney $(T_i, T_j)$ rejects $H_0$ for all pairs TrOC $T_i$, $T_j$ if $T_i$ and $T_j$ are siblings in the hierarchy.
4) Mann–Whitney $(T_i, T_j)$ rejects $H_0$ for all pairs of $T_i$, $T_j$ where $T_i$ is the immediate child of $T_j$.

In this step, we mined the constructs between treatments (with sequential nature) and a selected outcome. The output is a subset of treatments that are:

- Structurally different from each other considering the sequence of items, and/or
- Has a significant impact on the selected outcome.

## 4 APPLICATIONS

In this section, we study three software engineering applications of GWM. In GWM, analysts models the problem in the encoding phase and does not define the treatments. Instead, treatments and their relation with the outcome are retrieved from historical data. Hence, GWM is a method to extract insight from existing empirical data. We discuss three of these applications in depth.

**File editing patterns** - Subsection 4.1
**What?** For each file, we consider two status either being under edit or not. We consider the *edit time* as the total time that a file is being edited by a developer. Similarly, the *idle time* is the total time that a file is not changed. Considering idle and edit time, we applied GWM to mine if the sequence of the edit and idle intervals has a significant impact on the bugginess of files.
**How?** We first replicated a study [54] on a sample of 22 open source software products with 176,487 files in total. Then we applied GWM for mining patterns of file editing. Finally, we compared the results of GWM with the replicated study to show the added value of GWM.

---

> **Cause-effect construct:** sequence of the edit and idle intervals has a significant impact on the number of bugs in a file.
> **Outcome:** number of bugs in a file.
> **Encoding:** C (extended file idle intervals), and D (shorter idle intervals).
> **Treatments:** $C*, D*, CD*, D*C, (CD)*, C*D, C*D*, (C*D)*, (DC*)*, (C*D*)*$.
> **TrOC:** $C*$ (consecutive long idle intervals) has significantly higher number of bugs in a file compared to the $(C*D*)*$.

---

**Code ownership patterns** - Subsection 4.2:
**What?** Code ownership has been studied from different perspectives [7], [25], [54] to find the best way to decide who is most competent to receive a change request. Many models have been made to help this, although the effect of the sequence of contributors in changing a file was not considered. We applied GWM to mine if the sequence of touching a file by major and minor contributors has significant impact on file bugginess.
**How?** We defined the major and minor contributors of 176,487 files over 22 open source projects and applied GWM by considering *the number of post-release bugs* as the outcome measure.

---

> **Cause-effect construct:** sequence of major and minor contributors touching a file has impact on the number of bugs in a file.
> **Outcome:** number of bugs in a file.
> **Encoding:** A (file owner), B (other developers committing to a file).
> **Treatments:** 17 treatments (all the nodes of the hierarchy in Figure 2 other than $(BA)*$ and $BA*$.
> **TrOC:** $(A*B*)*$ has significantly more number of bugs in a file compared to $A*B*$.

---

**Release cycle time patterns** - Subsection 4.3:
**What?** When looking into the evolution of a software product, iterative release decisions such as duration of release cycle (short cycle or long cycles) are of importance to plan for a software release. More recently, several questions have been raised in this context, questioning the trade-off between release duration, the effort needed, and type of changes in releases [1], [6]. Some empirical studies have been designed to observe and report the impact of release duration on specific products [20], [28].
**How?** For 6,003 apps from Google Play store, we mined the duration of releases. Then, we encoded the duration between two consecutive release dates into short, medium or long release cycle and applied GWM.

---

> **Cause-effect construct:** sequence of release cycle time has impact on the apps' rating.
> **Outcome:** Apps' rating.
> **Encoding:** S (short release cycle), M (medium release cycle), and L (long release cycle).
> **Treatments:** 73 treatments over $\Sigma = \{S, M, L\}$.
> **TrOC:** 7 treatment-effect constructs between $L*M*$, $L*$, $L*S*$, $M*$, $(M*S*)*$, $S*$, and $(S*M*L*)*$.

---

We discuss the applicability of GWM in three case studies. However, GWM is not limited to these applications. The Gandhi-Washington Method extracts the recurring sequences of events with regards to their effect on a context-specific factor. In a nutshell, GWM is applicable to:

1) Data which has sequence by nature. For example, time sequences, process sequences, or development sequences.
2) Data presented in a system alongside numerical factors which provide a measure of the system performance (outcome). For example, code healthiness or estimated effort of a task.

3) Cases where multiple instances of data are available. Mining TrOCs is meaningful when we compare multiple itemsets.

## 4.1 File Editing Patterns

We first replicate a study on file editing patterns performed by Zhang et al. [54] and subsequently apply GWM on the same dataset and compare the results. Furthermore, we demonstrated how we can get more information and automatically extract all the treatment-outcome constructs using GWM. Following the approach and keywords reported by Ray et al. [38], we retrieved the number of commits related to bug fixes in each file. The extraction of bug related commits was done manually by two software engineers and the conflicts were resolved by one of the authors.

### 4.1.1　Case study data

We used the data of 22 trending projects from GitHub in the category of programming languages[3]. These projects had 176,487 files, 1,773 releases, and 4,501 developers in total. For each file we, mined the number of post-release bugs per file from git logs by considering the release and commit dates. Considering a release $r$, the number of post-release bugs is defined as the number of bugs occurring during the time between release $r$ and its consecutive release $r+1$.

We also estimated the edit and idle time for each file in these projects. Zhang et al. [54] considered the time span for editing a file as *edit time* and the time interval where no one was editing the file as the *idle time*. Zhang et al [54] studied Mylyn project which gave the actual value for edit and idle intervals. However, not having access to that data we used open source projects and estimated these time intervals.

We used Algorithm 2 to estimate the edit and idle time intervals for the files of our GitHub projects. We made the assumption that if a developer commits changes only to File I and File II at Datetime 1 and made changes to File III at Datetime 2, then edit time of File III is defined as $Datetime\ 2 - Datetime\ 1$.

---

**Algorithm 2:** Edit and idle time calculation

```
1  //datetime is the date and time of a commit.
2  for (each file) do
3      Ignore the commit that creates the file;
4      for (each following commit) do
5          Set commit's datetime as (A);
6          Set the datetime of the most recent previous
             commit, anywhere in the project, of this
             developer as (B);
7          edit time = (A) - (B);
8          (P) is the next developer committed to the file;
9          Set the datetime of the most recent previous
             commit, anywhere in the project, of developer
             (P) as (Y);
10         if (Y)<(A) then
11             conflict time = (A) - (Y);
12             idle time = 0;
13         else
14             idle time = (Y) - (A);
15         end
16     end
17 end
```

[3]. https://github.com/showcases/programming-languages

Note that we use this case study to demonstrate the added value of GWM and the contributions of this paper is not the actual and precise results of the Zhang et al. [54] case study and the estimated values are calculated to this end. In what follows, we go through different GWM usage scenarios using our GitHub dataset.

### 4.1.2　Replication of the study of Zhang et al.

Previous research by Zhang et al. [54] investigated file editing patterns. They hypothesized the existence of four file editing patterns namely concurrent editing, parallel editing, extended editing, and interrupted editing patterns. Patterns are the sequence of changing a file. These patterns were then analyzed in relation to the number of developers, the number of edited files, and the edit and idle time intervals. To demonstrate the benefit of GWM we only discuss interrupted editing patterns likewise one can apply it to extended patterns.

In this study, the output of the synthesis step is equivalent to what has been called as *pattern* by Zhang et al. [54]. To replicate their study, we calculated the maximum idle intervals of each file (IdleTime). Following Zhang et al. [54] a file follows the interrupted editing pattern "if and only if its IdleTime is greater than the third quartile of all IdleTime values". We calculated *IdleTime* of files and mined TrOCs.

The case study by Zhang et al. [54] reported that files following interrupted editing pattern are 2.0 times more likely to have future bugs. They used Fisher's exact test in conjunction with OR for their reasoning. In the same way, we found that files following the interrupted editing pattern are 1.86 times more likely to experience future bugs. With Fisher's exact test (p-value<0.001) and $OR = 1.86$. Our findings are aligned with the results reported in [54].

The study by Zhang et al. [54] examined predefined patterns and tested their likelihood of having future bugs. In other words, they assumed that specific treatments (sequence of idle or edit times) exist. Replicating their method but using GWM, we did not assume the existence of any treatment (idle or edit patterns), and we just encode the data as they proposed and GWM could mine the treatment-outcome construct. We compared the TrOCs with the patterns proposed by Zhang et al. [54]. Our findings are aligned with their reported results.

Now we go one step further. In the past study among all the idle and edit time intervals of a file, the decision was made based on the longest edit and idle interval for each file (i.e., the EditTime and IdleTime). Applying GWM, we discuss three different scenarios in the next subsections:

- We encode all the idle times for each file,
- We change the encoding for fine-grained analysis of idle times, and
- We extract patterns (treatments) from the idle and edit time instead of assuming the existence of two patterns (i.e., extended editing and interrupted edit- ing patterns) and extracting the TrOCs for testing the assumption.

### 4.1.3 Scenario 1: Synthesizing patterns

From applying GWM, we are interested in answering the following research question:

**RQ1:** Does any idle time sequence exist related to the bugginess of a file?
*Two sequences with significant impact on the number of post-release bugs exist and files having consecutive long (extended) idle interval have significantly more number of post-release bugs as compared to the rest of the files.*

**Encoding:** We encoded the time between a check-in and check-out of a file as idle time (see Figure 3). Having the idle time of all the files, we encoded sequences by using two letters.

*C:* Encodes the idle intervals greater than the third quartile value of all idle interval values (≥ 133.49 *hours*). This is called *extended idle time*.

*D:* Encode all the other idle intervals (less than the third quartile value).

In this model, each itemset represents a sequence of idle time intervals being longer than or equal to the third quartile values (encoded by C), or not (encoded by D). Comparing this with the Zhang et al. [54] model, we did not assume that there is an extended idle time pattern but we are looking to extract sequences of idle times which had a significant impact on the outcome (number of bugs).

**Categorizing:** We used the regular expression hierarchy over Σ = {C, D}. Using regular expressions, for example {C, C, C, D, D} is categorized as C*D* and {C, C, C, C} categorized as C*.

**Synthesizing:** We applied Algorithm 1 on the categorized itemsets and used the average number of post-release issues per file as the outcome. Using GWM, we mined two treatments having a significant impact on the number of post-release bugs.

- $C^*$: Consecutive long (extended) idle interval. 1.2% of files followed this sequence.
- $(C^*D^*)^*$: Combination of extended and not extended idle intervals. 98.8% of files followed this sequence.

The Boxplot distribution of the number of bugs (outcome) for each of the above patterns is shown in Figure 4 (a). $C^*$ had significantly more bugs.

**Discussion.** We considered all idle intervals for each file and applied GWM. The results showed that the consecutive extended idle intervals affect the average number of bugs significantly differently than other patterns. Also using

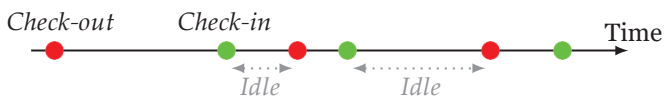

Fig. 3. Abstract model of idle time.

OR, we found that files that are edited with consecutive extended idle intervals are 2.3 times (*OR* = 2.3) more likely to experience future bugs in comparison to other files.

In addition to the results achieved by Zhang et al. [54], we found that the occurrence of extended idle intervals in each file affects the likelihood of a file's bugginess. *More specifically*, files having consecutive extended idle intervals are more defect prone ($C^*$). This might be because these parts of the code are not maintained for a certain period of time. Also, considering the turn over of team members this might occur because of the lack of knowledge about specific files. While GWM does not indicate causality, the analysis of the reasons could be the subject of other research.

### 4.1.4 Scenario 2: Change model granularity

To demonstrate how different encodings can solve different problems using GWM, we answer RQ2 using the same dataset as of Scenario #1:

**RQ2:** How does the results of more fine-grained encoding of idle time compare to the results of Scenario 1, considering the four categories instead of two for idle time intervals?
*GWM extracted five patterns. All these five sequences have a significantly different effect on the number of post-release bugs.*

**Encoding:** We want to extract more fine-grained idle time sequences as compared to the previous scenario. To acquire such sequences, we change the encoding of GWM and use an alphabet with four letters to encode idle time:

*C:* Encodes the idle time greater than or equal to the third quartile (133.49 hours, called *extended*) of all idle time values (like in Scenario 1).

*E:* Encodes the idle time less than or equal to the first quartile (≤13.24 hours, called *very short*) of all idle time values.

*F:* Encodes the idle time greater than the first quartile and less than or equal to the second quartile of all idle intervals (between 13.24 and 43.81 hours, called *short*).

*G:* Encodes the idle time greater than the second quartile and less than or equal to the third quartile (between 43.81 and 133.49 hours, called *long*).

In this model, an itemset represents the sequence of encoded idle intervals for a file.

**Categorizing:** We use the regular expressions over Σ = {C, E, F, G} to categorize itemsets. The items extracted from our dataset were categorized in 47 regular expressions.

**Synthesizing:** We applied Algorithm 1 on categorized itemsets, using the average number of post-release bugs as the outcome. The results show five patterns of idle time intervals with significant impact on number of post-release bugs:

- $C^*$: Consecutive extended idle intervals (same as Scenario 1-RQ1). 1.2% of all files followed this sequence.

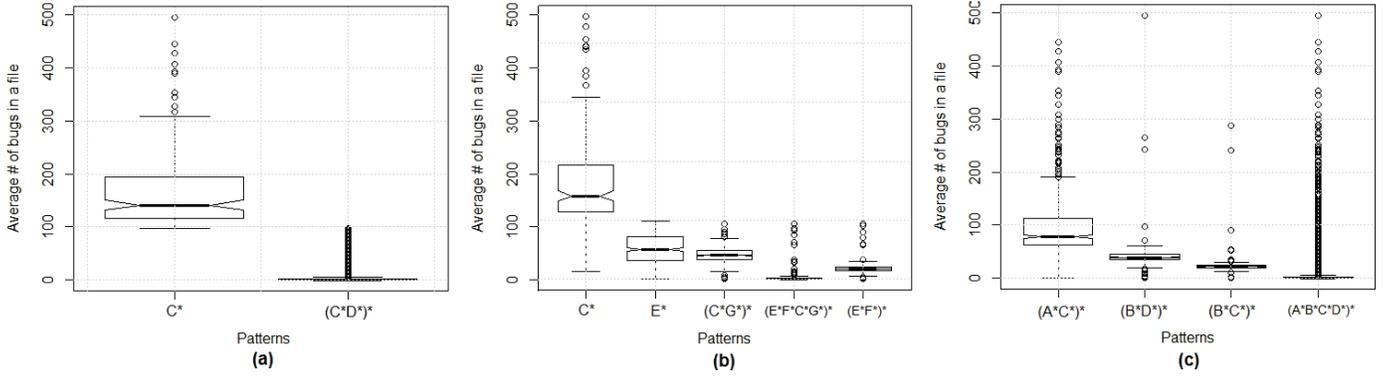

Fig. 4. Boxplot of average of bugs in a file for patterns of (a) idle intervals (b) fine grained idle intervals (c) idle and edit intervals in conjunction.

$E*$: Consecutive very short idle intervals. 1.7% of all files followed this sequence.

$(C*G*)*$: Combination of long extended idle intervals. 0.6 % of files in our case study followed this sequence.

$(E*F*)*$: Combination of short and very short idle intervals. 3.9% of all files followed this sequence.

$(C*E*F*G*)*$: Combination of idle intervals with different lengths. 92.6% of all files followed this sequence.

The box plot distribution of the number of post-release bugs (outcome) is shown in Figure 4−(b).

**Discussion:** Encoding enables us for flexible modeling of the problem. While in this scenario we followed the same model as Scenario 1, we retrieved more fine-grained sequences by defining more categories in the encoding step. We calculated the OR and Fisher's exact test values (as suggested by Zhang et al. [54]) and found that files that have consecutive long idle intervals are more than 4.201 likely to experience future bug compared to the files having consecutive short idle times. While comparing them with the mixture of all types of intervals i.e. $(C*E*F*G*)*$ they are 8.73 times more likely to have future bugs. On the other side, in comparison to the files with a combination of long and extended idle times, they are less likely to experience future bugs ($OR < 1$). Additionally, files with a combination of long and extended idle times are 9.1 times more likely to experience future bugs in comparison to files with a combination of idle times with differing lengths.

In comparison to Scenario 1, we gained more detailed and specified information about the treatments we extracted (previously $C*$). In addition, we can now capture and analyze all the idle time intervals, this led us into capturing treatments such as $(C*G*)*$ being 9.1 times more likely to experience future bugs.

It is worth to note that while the Mann–Whitney test result for $E*$ and $(C*G*)*$ was not significant (= 0.068), GWM keeps both as TrOCs because they relate to different branches of a regular expression hierarchy. In other words, the treatments were structurally different as these two treatments (sequences) are not siblings nor having parent-child relation with each other.

### 4.1.5   Scenario 3: Solve more complex problems

So far, we studied just the sequences of idle time for files. We are also interested in observing the sequences of edit and idle times of a file in conjunction and analyze the impact of these sequences on the files' bugginess. In this scenario, an itemset represents the sequence of both edit and idle intervals for a file. This model is shown in Figure 5.

**RQ3:** Does any idle and edit time exist that has an impact on the bugginess of a file?
*GWM extracted four sequences that are related to the average number of bugs significantly differently from each other.*

**Encoding:** We encode the items using an alphabet of four letters and follow the below encoding:

*A:* Encodes the edit time greater than or equal to the third quartile (≥ 42.39) of all edit time values.

*B:* Encodes all the other edit intervals (less than the third quartile of edit time).

*C:* Encodes the idle time greater than or equal to the third quartile (≥ 133.49) of all idle time values.

*D:* Encodes all the other idle intervals (less than third quartile).

Each itemset represents a sequence of edit and idle intervals for a file.

**Categorizing:** We used the regular expressions over Σ = {A, B, C, D} to categorize itemsets. Extracted itemsest were categorized in 59 regular expressions.

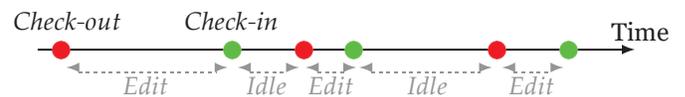

Fig. 5. GWM can encode idle and edit times together.

**Synthesizing:** By applying Algorithm 1 to the categorized items, we found four sequences that each of them has an impact on the number of bugs significantly different from each other:

- $(A*C*)*$: 0.4% of files follow extended edit intervals and extended idle intervals.
- $(B*C*)*$: 0.8% of files had the sequence of not-extended edit interval and extended idle time.
- $(B*C*D*)*$: 1.3% of files had the combination of idle interval types with not-extended edit time (all types of encoding other than A).
- $(A*B*C*D*)*$: 97.5% of files have a combination of different types of edit and idle intervals.

The box plot distribution of number of post-release bugs (outcome) is shown in Figure 4−(c).

**Discussion:** Considering the idle and edit intervals together and measuring the likelihood of future bug occurrence using Fisher's test and Odds Ratio (OR) (as suggested by Zhang et al. [54]) we found that files that were not edited for a long interval and were idle for a long period (i.e. $(B*C*)*$) are **8.6** times more likely to experience future bugs in comparison to files that have a combination of all types of edit and idle time intervals (i.e. $(A*B*C*D*)*$). Additionally, files that have not been edited in extended periods of time (i.e. $(B*C*D*)*$) are **1.97** times more likely to experience a future bug in comparison to the files following extended edit and idle interval sequences (i.e. $(A*C*)*$). Files that follow $(B*C*D*)*$ are **7.1** times more likely to experience future bugs in comparison to the files following $(A*B*C*D*)*$.

## 4.2 Code ownership sequences

The writer (developer) of a line of code has the most knowledge about that part of the code. The owner of a piece of code (for example a file) is determined as the developer who wrote the most lines in that piece of code.

Code ownership is defined as the percentage of lines of code that a developer owns in a file [14]. Code ownership was studied in many different ways in order to find the most competent and knowledgeable person to change the code [7], [14], [25]. We analyzed the relationship between code ownership sequences and the number of bugs related to that file. We studied the following research question:

**RQ4:** *Do any code ownership sequences exist related to the number of bugs for a code file? Applying the Gandhi-Washington Method on a large scale dataset from GitHub, we found two sequences of code ownership in a file that have a statistically significant relationship with the healthiness of a program file.*

**Encoding:** Analyzing code ownership for each file, we considered two types of contributors; the contributor who owns the file (she has the most code churns in that file) and the rest of the developers who committed changes to the file. Using $\Sigma = \{A, B\}$:

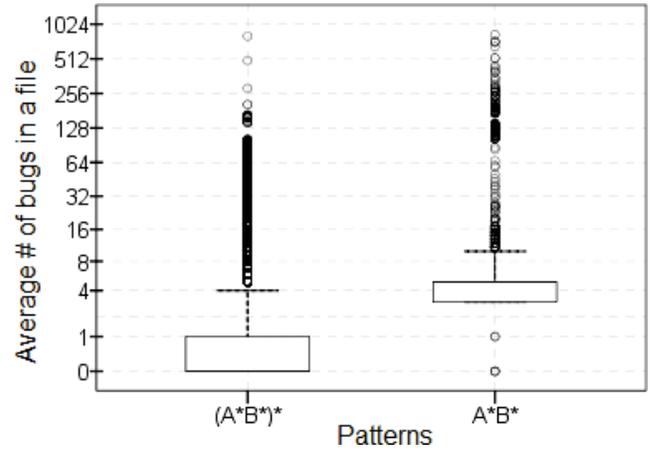

Fig. 6. Boxplot of the # of bugs for code ownership sequences among 175,995 files from GitHub.

*A:* For each file, $'A'$ represents the *file owner*. For each file, we calculated the code churn (number of changed lines of code) per developer during the lifecycle of the file. We considered the person with the highest number of churn during files' life cycle as the owner.

*B:* $'B'$ represents the *rest of the contributors to any given file*. These developers contributed changes to a file but they do not own the most lines of code in the file during its life cycle.

The owner of a file could be $'A'$ or $'B'$. After we defined file owners, we assigned $'A'$ or $'B'$ to the owner for each release of the file. For every file in a project, we created an itemset within which the items show if the owner of the file overall is also the owner in each of the releases. In short, $'A'$ is the file owner that contributed the most lines of the code overall releases while $'B'$ contributed most lines of the code only in particular releases of a file.

For example, an itemset such as $\{A, A, B, B, B\}$ shows that $'A'$ created the file and initially committed changes while the most recent changes were by other developer(s). The file owner of this file ($'A'$) cumulatively made most changes to this file (note that $'B'$ is not one specific developer but the other developers not being the file owner). In this case study, we ended up with 175,995 itemsets as some of the files were never changed in their projects' life-cycles. We selected the average number of post-release bugs for a file as the outcome to analyze the relationship between the ownership sequences and the bugginess of the files.

**Abstraction:** We used the regular expression hierarchy over $\Sigma = \{A, B\}$ (like the one in Figure 2) to categorize the itemsets. This enumerated hierarchy has 19 nodes, of which 17 had encoded itemsets categorized into them. Using regular expressions $\{A, A, B, B, B\}$ is categorized as $A*B*$.

**Synthesis:** We applied Algorithm 1 on the categorized itemsets and used the average number of post-release bugs per file as the outcome. As the results of this process, we found two sequences (treatments) as $A*B*$ (22.3% of the files) and $(A*B*)*$ (77.7% of the files). Files with the ownership sequence $A*B*$ have significantly more post-release bugs as compared to the rest of the files. Figure 6 shows the distribution of the number of post-release bugs in the files with $A^*B^*$ and $(A^*B^*)^*$ ownership sequence. $A^*B^*$ represents files for which the developer who mainly owns the file in the project has the ownership of the file in early releases, but later the release code ownership of a file is transferred to another developer.

The result of this analysis shows that files which have ownership sequences of $A^*B^*$ have significantly more post-release bugs compared to other files. Significantly more bugs in $A^*B^*$ might be because of the unfamiliarity of later stage file owners ($B^*$) with the code that was maintained largely by the main owner for a while ($A^*$). Further investigations are needed to find the actual cause for this problem.

**Discussion:** Code ownership TrOCs are not targeting predictions, but rather determining if there is a statistically significant relationship between a sequence of encodings for owning a file and the number of post-release bugs. Bird et al. [7] reported that the number of low expertise owners has a relationship with both post-release and pre-release failures. They showed that higher level of ownership for the top contributor results in fewer failures. Their study considered the proportion and level of ownership, but the impact of sequences of ownership (as studied here) was not discussed. The results of our study showed that files following $A^*B^*$ sequence (consecutive edits by the file owner and later consecutive edits by other contributors) are 1.74 times more likely to experience future bugs (p-value of Fishers test = < 0.001 and $OR$ = 1.74). More advanced and dynamic ownership scenarios could be studied similarly. However, this would require a different encoding.

### 4.3 Release cycle sequences

A product manager is concerned about the release date for next version of a software product. She is following a fixed cycle schedule to release a version of her code every two weeks. Now she is facing a major bug in the code. Fixing the bug would delay the version release. She decides to release the product as she believes that users are expecting to receive the new version based on the scheduled fix cycles. In other words, she assumes that following a fixed release cycle increases customer satisfaction. GWM can find the relation between fixed, short release cycles and customer satisfaction.

Defining release cycle strategies for a software product are a challenge for many practitioners and researchers [1], [20]. Traditional software development processes were changed with the introduction of iterative processes. Later, agile practices enhanced the development process a step further by using short release cycles. The effect of short and long release cycles on different aspects of software products like teams productivity, requirements engineering and specification were discussed.

We have always discussed the fixed short release cycles (agile development with short scrums) versus the long release cycles (traditional iterative processes). Mapping this terminology into GWM, we represent short release cycles by $S$ and long release cycles by $L$. Analyzing a series of subsequent short release cycles is then represented as $S^*$, and the analysis of consequent long release cycles is represented as $L^*$. Having this in mind, the GWM contributes in two ways:

1) It discovers more in-depth release strategies among different software. In this way, we can extend the discussion of traditional versus agile iterations to include more diverse release strategies such as long release cycles followed by a consequent short cycles (for example for stabilizing a major release).

2) It statistically analyzes the effect of these sequences (treatments) on a context-specific performance measure (outcome). For example, the effect of consequent short cycles, $S^*$, on the number of bugs.

**RQ5:** Do any release cycle time sequence exist with significant impact on the mobile apps' rating? *Applying GWM we mined seven sequences of release cycle time that has significant impact on the app rating.*

Release cycle time and its variation matters a lot in mobile app development [31], [32], [33], [34], [35]. To answer RQ5, we analyzed 6,003 apps from Google Play and we selected the app rating as the outcome. The apps in this sample have different numbers of releases, ranging from 3 to 186. These 6,003 apps had 60,588 releases in total. Each app has a rating between zero and five which is the average rating granted by app users.

The release cycle is the time between two consequent releases. In this sample, the release cycle duration follow a power-law like distribution. In this data, one quarter of the release cycles are less than five days, while one quarter takes more than a month. Applying GWM to a set of 6,003 apps, we found seven particular release cycle time TrOCs.

**Encoding:** Having release cycles of each app as items in the itemset, we applied frequency based distribution on the release cycles. This resulted in three clusters called $S$(short), $M$ (medium), and $L$ (long).

$S$:    represents cycles between 1 to 7 days of duration (less than a week),

$M$:    represents cycles between 8 to 21 days (up to three weeks) and,

TABLE 2
Release cycle TrOCs

| ID | Treatment | Description | % of occurrence |
|---|---|---|---|
| P1 | L*M* | Combination of $L$ and then $M$ cycle types. | 14.1% |
| P2 | L* | Subsequent release cycles of [23, 1365) days. | 23.7% |
| P3 | L*S* | Combination of $L$ and then $S$ cycle types. | 9.5% |
| P4 | $(S^*M^*L^*)^*$ | Combination of all cycle types. | 35.1% |
| P5 | M* | Subsequent release cycles of [7,23) days. | 3.5% |
| P6 | $(M^*S^*)^*$ | Combination of $M$ and $S$ cycle type. | 7% |
| P7 | S* | Subsequent release cycles of [0, 6) days. | 7.1% |

*L*:     represents cycles 22 days or more (i.e., more than three weeks).

For example, {20, 16, 19, 15, 13, 21, 20} and {21, 122, 81, 61} represents release cycle times for two different apps where the first one is encoded to MMMMMMM and the second itemset is encoded as MLLL. Encoding in GWM is flexible and can be customized in any context. Alternatively, in this case, encoding could be different using experts defined ranges or clusters defined by using the k- means algorithm [26].

**Abstraction:** A regular expression hierarchy over *S*, *M*, *L* with 130 nodes was used. Categorizing the 6,003 release cycle sequences, 73 nodes of this hierarchy were filled with at least one itemset. Using the regular expressions, *MMMMMMM* is categorized in $M^*$ and *MLLL* categorized in $ML^*$.

**Synthesis:** We applied Algorithm 1 on the 73 nodes of the regular expression hierarchy. Using app rating as the outcome, we obtained seven unique release sequences that significantly affect the rating of the apps within our sample (seven TrOCs). Table 2 and Figure 7 show the final treatments extracted by GWM and their outcome distribution. Apps that have consecutive long release cycles (more than 3 weeks) followed by consecutive short cycles (less than a week) have the highest median of app rating ($L^*S^*$), followed by apps that have consecutive long and medium release cycles ($L^*M^*$). The lowest median app ratings correspond to apps with long release cycles exclusively ($L^*$).

In Appendix I, for a subset of this data, we provided a detailed description of performing all the synthesis steps.

## 5 GWM TOOL SUPPORT

We provided a tool support for the proposed method. The GWM support tool is a `Windows-based` program which has been tested on `Python 3.5.2`, though it may work on earlier versions. The tool has a graphical user interface (GUI) supported by `PyQt` as well as a shell access to support users of all levels.

The tool has a separate interface for each of the three phases of GWM as well as a dedicated interface for analyzing the fitness of statistical tests to support in-depth analysis and inferences by running multiple statistical tests:

**Encoding:** The encoding step translates lines of data from a `csv` file and outputs a string of characters representing an itemset. If the input data is non-categorical, the tool provides support for expert-based discretization and frequency-based discretization. The output of this phase is a `csv` file that automatically transforms as the input for abstraction.

**Abstraction:** The abstraction step matches encoded strings into enumerated regular expressions. The results of the process are given in the Categorized Strings window and shows that which itemset was categorized in which regular expression. The output is a `csv` file which is used as the input for synthesis phase.

**Synthesis:** Synthesis phase merges separate regular expressions where the Gandhi Washington Factors categorized into them are statistically different using Mann-Whitney U-test as a default (that could be alternated). The output is provided in the format of a table showing the treatments, mean of the group and the sample size (number of itemsets) for each treatment. The distribution of the outcome is visualized trough box plots. The Mann-Whitney is a non-parametric test applicable also for small sample sizes. The only assumption for the test is that the distributions of the two groups are the same under the null hypothesis. The distribution of the groups is checked within the tool (see Figure 8 - (D)).

**Fitness of statistical tests:** Every two treatments can be selected and their constructs can be compared through a series of statistical tests and visualizations. The format of the input file is the same as in the synthesizing and so is useful to verify certain results.

In Figure 8, we provided some screenshots of the tool. A test generator tool is available along with the Gandhi-Washington tool. The test generator will help to run controlled experiments with the tool as well as logging the time and determinism. The tool supports up to five encoded items. However, the design is extensible.

GWM implementation, test generator and described test cases in Section 7 are all accessible via our website[4].

## 6 RELATED WORK

In *software engineering*, patterns are used to "encapsulate knowledge for constructing successful solutions to recurring problems" [39]. Patterns have been used on:

- Code-level constructs: such as program generation, re-usability, and code defects,
- Design level constructs: such as design skills, and architecture design, or
- Knowledge and communication: such as work experience, pattern organization, and documentation [39].

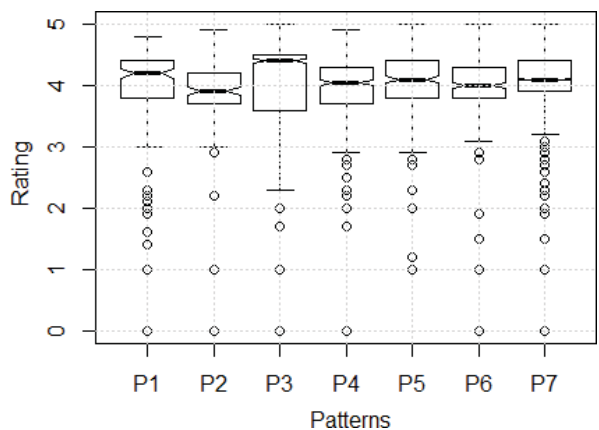

Fig. 7. Boxplot of apps rating for release cycle time TrOCs among 6,003 Android apps.

Pattern selection, application, and modularity of patterns in software engineering were evaluated in most cases by domain experts, researchers, or participants in an empirical study [39].

---
4. http://ucalgary.ca/mnayebi/tools-and-data-sets

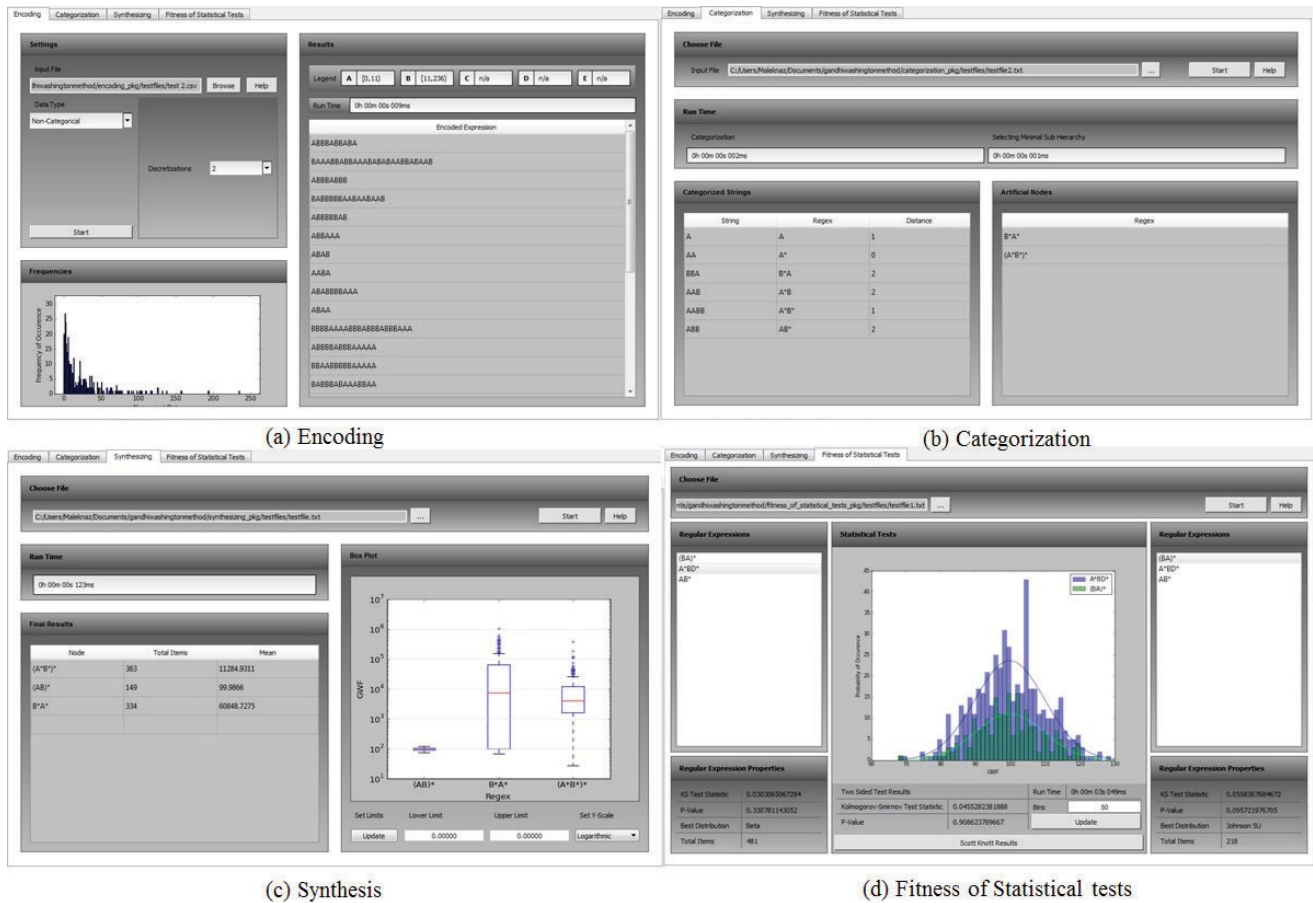

Fig. 8. The GUI screen-shot of the Gandhi-Washington tool support. The tool supports all the three phases of the GWM process and provides additional support for verification of the results.

Xie et al. [52] and Hassan and Xie [17] defined three broad categories for software engineering data, naming *sequences*, *graphs*, and *text*. We focused on sequential data in this paper. Xie et al. introduced "execution traces collected at run-time, static traces extracted from source code, and co-changed code locations" as the most prominent examples of sequential software engineering data which are mainly related to programming, bug detection, maintenance, and debugging tasks. Software development deals with a large amount of data created from sequential activities, events, and decisions. One of the foreseen challenges for mining sequential data was the complexity of data and mined patterns [52].

In the field of data mining, sequential pattern mining discovers frequent patterns in a database. Sequential pattern mining approaches targets databases with sequences of ordered events with or without the concrete notion of time [24] such as the sequence of customer's transaction in an online store. A wide range of applications for these approaches has been discovered from web-access patterns

TABLE 3
Comparison of pattern mining approaches.

| Approach | Solution | Confinement | Instances |
|---|---|---|---|
| Apriori-based | Construct all the possible sequences by generating the candidate sequences and build patterns one item at a time iteratively and traverse the search space. | Any sub-pattern of a frequent pattern must be frequent as the measure for pattern interestingness. | Apriori [2] [12] SPADE [53] SPAM [4] GSP [46] |
| Pattern-growth-based | Search of patterns in a specific part of a given database by making the suffix and prefix trees of data. These algorithms do sequence pruning in order to prune candidate sequences early in the process. | Based on sampling and compression. Looking into frequency and periodicity as the measure for pattern interestingness. | FreeSpan [15] PrefixSpan [16] SPIRIT [13] |
| Temporal patterns | Search for the patterns of interactions in a time-ordered input sequence. | Looking into frequency, length, and periodicity as the measure for pattern interestingness. | Time-based methods [18] [51] |

TABLE 4
Comparison of temporal pattern mining and Gandhi-Washington Method.

| Mining temporal patterns | Gandhi-Washington Method |
|---|---|
| Consists of events with no natural points indicating the start or stop of an event. | Start and stop of events are defined to be consecutive version releases i.e., the release cycles. |
| Use event-folding technique (sliding time window) to partition sequence of event. | Natural sequence of events are defined by releasing a new product into the market. |
| Create additional candidate episodes from subset of maximal episodes (top-down approach; from genera to more specific). | Create artificial nodes (bottom-up from the most specific to the more general regular expressions). |
| Use frequency, preciosity and length of patterns for interestingness. | Use of statistical inferences on a context-specific factor (outcome). |
| Looking for patterns that follow interval sequences as a pair of (timestamp, event). | Extracting patterns of timestamps and characterize them by events. |

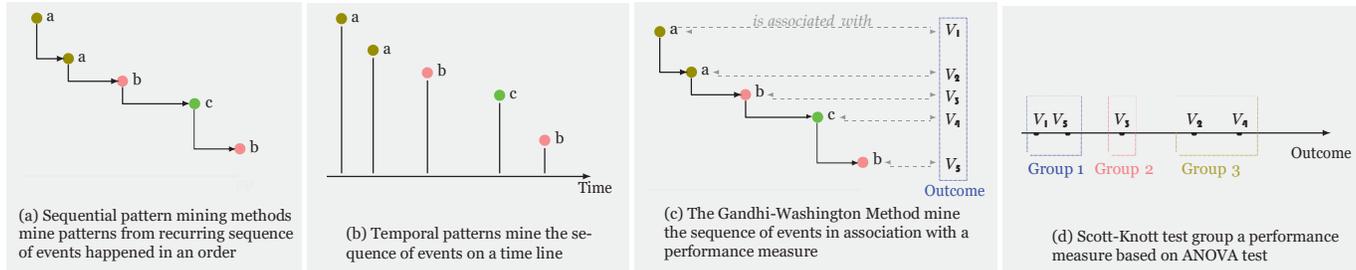

(a) Sequential pattern mining methods mine patterns from recurring sequence of events happened in an order

(b) Temporal patterns mine the sequence of events on a time line

(c) The Gandhi-Washington Method mine the sequence of events in association with a performance measure

(d) Scott-Knott test group a performance measure based on ANOVA test

Fig. 9. Comparison of the input data for sequential pattern mining, temporal pattern mining, Gandhi-Washington Method and Scott-Knott approach. In the figure a, b, c are the events and $V_1$ to $V_5$ are the values of the impact factor. (a) the output of the sequential pattern mining methods are itemsets that appeared in the data with certain frequencies (b) the output of temporal pattern mining methods is a set of patterns defined based on frequency, length and periodicity of patterns, (c) GWM finds patterns that are related to a performance factor, (d) Scott-Knott test forms clusters with statistically significant difference in the group means.

to the analysis of DNA. Several solutions were proposed to efficiently mine sequential patterns and several comprehensive and comparative analysis of these approaches exist [24], [29], [37]. In Table 3 we summarized the three main methodologies and the instance of approaches that follow each method.

We compare and differentiate Gandhi-Washington Method with existing sequential pattern mining methods and clustering methods based on group means. Different algorithms for mining sequential data reflect different levels of information, and the choice of algorithm depends on the tasks mining requirements [52].

### 6.1 Comparison of GWM with sequential pattern mining methods

Sequential pattern mining approaches are mainly categorized as Apriori-based [2] and pattern-growth methods [15]. These methods use frequency (in terms of user-specified minimum support thresholds) and length as the measure for pattern interestingness. In one instance of sequential pattern mining approaches regular expressions are used [13] which enable users to express the specific category of sequential patterns that are of interest to them. The solution and constraints for all these approaches are compared in Table 3.

Song et al. [45] used Apriori-based algorithms to mine methods to predict defect associations and defect correction effort. Michail also used Apriori-based algorithms to mine patterns in code library usage [27]. Pattern growth-based algorithms were used by Lo et al. [22] to mine software behavioral specification.

Gandhi-Washington Method analyzes the hierarchical relation between events by using regular expressions and merges the sequences when it cannot find statistical differences between them, while previous approaches used frequency, length, and periodicity as the factor to select patterns. With this aim, these approaches are fundamentally different while both target sequential and ordered itemsets. In Figure 9 we demonstrate the difference between sequential pattern mining methods and GWM.

The most well-known sequence is defined in the order of *time*. The Mining temporal sequences for discovering patterns [18] is searching for the patterns of interactions in a time ordered input sequence. This approach is looking into frequency, length, and periodicity of patterns as the measure for interestingness. First, it partitions the event sequence into the maximal episodes and creates an initial set of candidates from these episodes. Second, the approach is generating additional candidate episodes as a subset of maximal episodes and evaluates the interesting episodes by computing the compression ratio to select the interesting episodes as the final patterns. The comparison between temporal pattern mining and the Gandhi-Washington Method is presented in Table 4.

Wasylkowski and Zeller [49] used temporal patterns to mine violations of operational preconditions in code. Also, Herzig and Zeller [19] mined temporal process patterns that encode key features of the software process and validate them automatically. Uddin et al. [48] proposed a method for mining temporal API patterns to detect API usage patterns in terms of their time of introduction into client programs.

## 6.2 Comparison of GWM with clustering methods

Using the analysis of variance to split treatments into homogeneous sets [42] formed a class of clustering methods. This specific class of clustering algorithms introduced and known as Scott-Knott test [42] which uses group means to partition the data.

Scott-Knott is a hierarchical clustering algorithm that creates non-overlapping groups of treatments using the ANOVA test. The method of Scott-Knott uses a hierarchical and divisive clustering method that impose a hierarchy of clusters based on group means. In each step, the best clustering is selected by the sum of squares within groups. Scott-Knott's termination criterion is based on results of ANOVA test at each stage of the procedure which was corrected for Type I error [8]. Tian et al. [47] used Scott-Knott test to rank feature importance in characterizing high rated mobile apps.

In comparison to Scott-Knott, the Gandhi-Washington Method considers the sequence of event occurrences along with the distribution of treatments and hence is different during the synthesize. This makes GWM suitable for analyzing the impact of processes and decision sequences on the software metrics. This is visualized in Figure 9. We compared the input data of different methods mentioned above in Figure 9 and in Table 3.

## 7 DISCUSSION

In this section, we discuss some further aspects of GWM's implementation. We look at possible solutions for more complex encoding for mining treatment-outcome and we will show that GWM is deterministic. In addition, we discuss the scalability and complexity of the method.

### 7.1 Multiple encoding

Encoding is flexible and in the case where multiple and related variables exist, multiple encoding is possible. In the *Alice, Carol, and Bob* example (Section 3.1), the *lines of code (LOC)* in conjunction with the number of bugs each of them reported is of interest. In this case, multiple encoding is applied. We discretize the LOC into two groups naming 'A' for LOC below 5,000 and 'B' for LOC equal to or above 5,000. Each of Alice, Bob and Carol might be in one of the groups below.

| Bug group | LOC group | Encode Item |
|-----------|-----------|-------------|
| L | A | U |
| M | A | V |
| H | A | W |
| L | B | X |
| M | B | Y |
| H | B | Z |

In this way, the encoded itemset for Alice, Carol, and Bob sequence might be such as 'UWX'. Once an analyst models the problem and encodes the events, GWM abstraction, and synthesis steps automatically retrieve TrOCs. The abstraction and synthesis steps do not need the analyst's participation. It is therefore easy for an analyst to try different encodings and find the proper match.

### 7.2 Defining abstraction hierarchies

The difficulty of extracting regular expressions has been acknowledged in other software engineering papers [9], [10]. Among all possible regular expressions and hierarchies, we defined a specific hierarchy by considering regular expressions from using all letters from Σ at most once and by considering the permutation of letters, brackets from applying Kleene star.

While the number of regular expressions is infinite, we fore see two problems occurring with too detailed hierarchies:

1) With more detailed regular expressions, the number of sequences following that regular expression would decrease. Hence, the number of samples for running statistical tests would be at risk to be not applicable. As a result, the node would be merged into a more coarse-grained node.
2) The patterns become less intuitive and interpretable.

Considering the context and setup of the examination, an analyst can check if more detailed regular expression can help in the analysis. The "tree generation abstraction algorithm in our GMM tool can be extend to accommodate more detailed hierarchies. That way, the algorithm can be extended to generate more detailed hierarchies for the abstraction phase. In addition, any desired hierarchy of regular expressions can be added as a `.csv` or `.txt` file to the "tree generation module to be used for abstracting the sequences.

### 7.3 Fitness of statistical tests in the synthesis phase

When multiple options are possible during the synthesize, we first use the effect size to decide on merging. We used the Odds Ratio (OR) [43] measure (as suggested by Zhang et al. [54]). However, if the effect size for a node, its parents and its siblings are very close ($\Delta OR < 1$) we consider OR as non-decisive. In these cases, we use the p-value of the Mann–Whitney test to make the decision.

A higher p-value shows a weaker evidence against the null hypothesis. Consequently, if a node has more than one parent and the synthesis has decided to merge it, the node will be merged with the parent that returns the higher p-value in a pairwise Mann–Whitney test. In cases of ambiguity in selecting one node among all its siblings, the node that rejects the null hypothesis least frequently will be selected. If there is a draw between the siblings for the fewest rejected null hypotheses (as they reject the test same number of times), then the one with the highest p-value will be merged first. Other effect size measures can alternatively be used.

The Mann–Whitney test in synthesis is applicable to ordinal outcomes. The test could be replaced by other statistical tests in case of non-ordinal outcomes. If we apply the Mann–Whitney test on two groups of itemsets with a non-identical distribution of outcome, the test compares the mean rank between the groups. Replacing the Mann–Whitney test with tests such as the t-test or even group tests such as ANOVA or Kruskal-Wallis is possible, although one needs to adjust the interpretation of TrOCs in each case. Furthermore, we adjusted p-values by using Bonferroni error correction for multiple comparisons.

## 7.4 Determinism and complexity

The main computational complexity of GWM comes from the abstraction and synthesis steps as the encoding is done by the analyst. In order to exhibit the performance of GWM in practice and later to support the applications, we implemented GWM using Python. In this section, we report the computational complexity and actual run-time of an implementation of abstraction and synthesis. The run-time is measured using an `Intel CORE i7-260M` CPU. The results show the scalability of this method.

We also examined determinism of the abstraction and synthesis. We used syntax-based coverage criteria [3] and enumerated regular expressions to generate test cases. We developed a test generator tool for automated testing of GWM. For testing the abstraction, the testing tool checked the strings to see if they were categorized back into the regular expression that generated them. To test the synthesis, we assigned outcomes to the strings in a way to retrieve specific regular expressions as a TrOC. We then compared the results with our expected outcomes.

**Abstraction:** Abstraction consists of two sub-processes, (i) reaching nodes in the regular expression hierarchy and (ii) sequence matching (analyze if the string is producible by the regular expression). Scanning all of the regular expressions to find a suitable node has a complexity of $O(n)$, with $n$ being the number of nodes in a regular expression hierarchy. While visiting each node in this hierarchy, the string is analyzed against the regular expression stored in the node. To do so, a finite state automaton based search is used with a computational complexity of $O(|\Sigma| \, m)$, $m$ being the number of strings processed by the abstraction process and $\Sigma$ being the size of the input alphabet $|\Sigma|$. Hence, the complexity of this process is $O(|\Sigma| \, nm)$.

To validate abstraction, a series of strings was produced by a generator tool. The generator tool takes each enumerated regular expression and produces strings using rules of formation. These strings were used as the input for our implemented abstraction. We tested how many strings were classified in nodes equal to their generator regular ex- pressions. 10,000 sample itemsets were generated from 500 regular expressions by using the test generator. The output from the abstraction mapped one-to-one to the generated test cases and was 100% accurate and deterministic. All of the generated strings were categorized in the regular expressions that they were produced from (accuracy), which was the most specific regular expression they could match to. We ran each test case multiple times and the results were consistent in all the runs.

To demonstrate the scalability and performance of the abstraction, we created five different sets of 500 strings. Each bin of 500 strings had the same length from 1 to 16 letters. The time performance of the synthesis over this dataset is illustrated in Figure 10.

**Synthesis:** Following Algorithm 1, a graph traversal approach is employed to apply a series of Mann–Whitney tests between nodes in the hierarchy. Using Depth First Search (DFS) [11] and including pairwise comparisons, the algorithm has the time complexity of $O(n(m + n))$, with $m$ being the number of edges and $n$ the number of nodes in the hierarchy graph. Examining nodes for merging needs another graph traversal and updates to some of the test results in case a node is merged. This has the time complexity of $O(n(m + n))$. To test the correctness of the synthesis we tested two conditions,

(i) If the synthesis merges nodes with similar outcome distributions, and
(ii) If the synthesis returns nodes that are significantly different in terms of outcome.

We generated 100 strings per regular expression over four different alphabets $\{A, B\}$, $\{A, B, C\}$, $\{A, B\}, \{C, D\}$, and $\{A, B, C, D, E\}$. In order to check (i) the following partitioning strategies were used:

- We assigned a random outcome value $x$ to all the strings belonging to half of the regular expressions and the random outcome $y$ to the rest, where $x = y$. We expected to have two final treatments.
- We assigned a random outcome value $x$ to the strings belonged to one third of regular expressions, one third a random outcome $y$, and one third a random value $z$, where $x \neq y = z$. We expected to have three final treatments.
- We assigned a random outcome $x$ to all the strings belonged to one fourth of the regular expressions, one fourth a random outcome $y$, one fourth a random value $z$, and one forth a random value $w$ where $x = y = z = w$. We expected to have four final treatments.

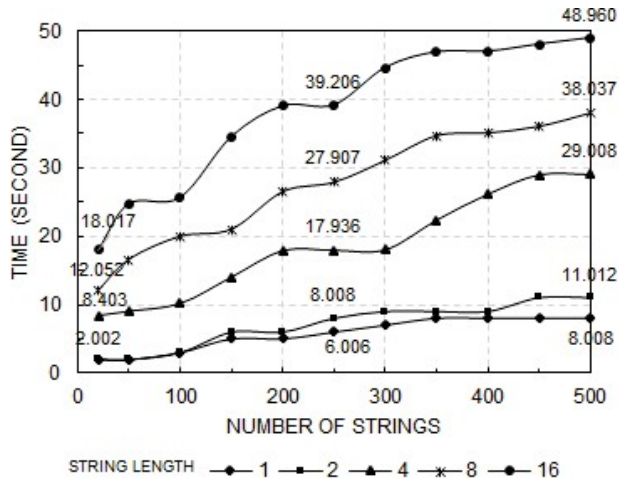 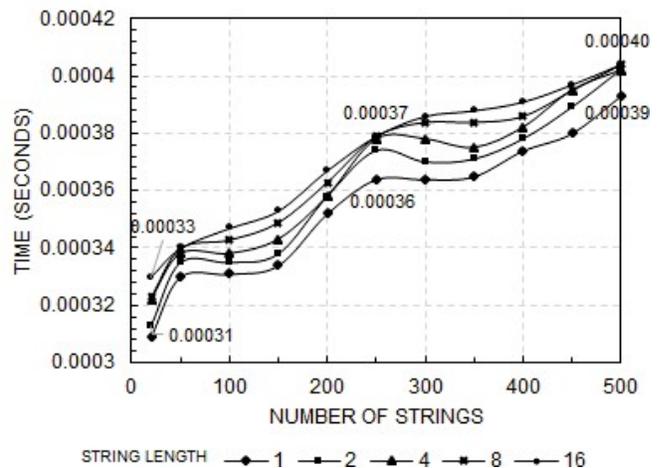

Fig. 10. Time performance of the abstraction (on left) and synthesis (on right) for 2500 cases. Each line shows the performance for 500 strings of the same length.

We ran each partitioning strategy on four different alphabets. We observed over 12 runs of synthesis that all the strings were merged as expected (100% accuracy).

To test condition (ii), in each test we targeted specific regular expression to be retrieved by GWM as the TRoC. For the enumerated regular expressions over different alphabets, a random node was selected. Random outcomes were assigned to the strings generated by the selected regular expression. We assigned outcome with random value x to all the other strings belonging to the rest of the regular expressions (making them merge together). We ran 20 tests (five tests for each alphabet set). For all the tests, GWM returned the targeted regular expression as TRoC and matched with our expected outcome (100% accurate). We ran each test case multiple times and the results were consistent in all the runs. To demonstrate the scalability and performance of the synthesis, we created five different sets of 500 strings. Each bin of 500 strings had the same length from one to 16 letters and had random outcomes. The time performance of the synthesis over this dataset is illustrated in Figure 10.

### 7.5 Comparison between GWM and Apriori

The fundamental difference between Apriori and GWM is that the order of items within an itemset is not essential for Apriori. In contrast, the sequence of items is the main concern and motivation for GWM. Also, GWM studies itemset in conjunction with a performance measure, which is not the case for Apriori.

We applied Apriori to mine frequent itemsets for the "code ownership" application in Section 4.2. We used the value of 0.7 as the level of minimum support. Maximum length was set to three for the same purpose. Below are the mined frequent itemsets determined that way. Table 5 shows the patterns of item size 2 and 3 that have been retrieved by applying Apriori to the former data. We can see that the results are rather different in its nature: There is no abstraction in Apriori, and the results have no interpretation in terms of a performance measure. In summary, we have different questions being answered by the two methods:

Apriori: What are frequent code ownership patterns of a file for different item sizes?

TABLE 5
Frequent itemsets mined for code ownership sequences of Section 4.2.

| Item size 2 | Item size 3 |
|---|---|
| {A, A} | {A, A, A} |
| {B, B} | {B, B, B} |
| {A, B} | {A, A, B} |
|  | {A, B, B} |

GWM: What are the code ownership patterns that impact the number of bugs and that are statistically different in the performance measure?

## 8 LIMITATIONS

GWM facilitates the structuring and packaging of knowledge gained from empirical investigations. However, there are various limiting factors for its applicability:

**Scope:** GWM is applicable for empirical studies that are concerned with understanding the impact of structure and impact of sequences of activities and events on a performance measure. If a study does not target mining and understanding sequential data, then GWM is not applicable.

**Granularity:** The idea of GWM is finding commonality in sequences and their outcome. The definition of structural commonality as done in abstraction imposes abstraction from details in the sense that sequences {A, A, A, A, A, A, B} and {A, A, B} are considered the same (expressed as $A*B$).

**Cognitive limitation of modeling:** The encoding phase of GWM needs human experts to define items in a way to model the problem. Different modelings of the items would solve different problems. Correct modeling for solving the right problem is essential in GWM. Encoding is analysts' responsibility and human error is unavoidable which may pose a threat to the validity of the results. It is possible that the change in the encoding implies changes the results of GWM, as it is true for the change in parameter settings of any analytical investigation. While this analysis can be done by changing the corresponding parameters, it is not part of the original GWM method.

**Run-time and algorithmic complexity:** So far we discussed the run-time and complexity of the method using up to five categories for encoding expecting that most problems can be addressed with a small alphabet. The relation between regular expressions makes it hard to define the enumerated hierarchy of items. It is expected that more computational resources are needed for the higher number of categories.

## 9 SUMMARY

The Gandhi-Washington Method is an approach to analyze the sequence of recurring events and items studied in relation to a context-specific performance measure. GWM represents a structural and unified method to determine the effect of different software engineering decisions and event sequences on projects, processes and products performance. GWM combines the use of regular expressions with the application of statistical tests. The encoding phase provides a flexible means for analysts to model the problem using a set of alphabets. Abstraction and synthesis are automated steps in GWM which condense the data and later aggregate sequences of items based on their commonality in structure and their effect on a software performance measure.

GWM shows which recurring sequences in software processes do significantly affect performance measures and retrieve them as TrOCs. From the potentially broad range of applications, we demonstrated the usefulness of GWM on code editing, code ownership, and release cycle time analysis. Like most statistical methods, GWM is not intended to claim causality, as confounding factors cannot easily be excluded and the patterns retrieved by GWM are tied and limited to the context-specific performance measure and changing this factor in the process of GWM may result in different patterns. Also, further analysis is needed to measure the sensitivity and robustness of the results gained in dependence of the underlying datasets.

## ACKNOWLEDGMENTS

We wish to thank Enoch Tsang, Kurtis Jantzen, Samarth Sinha, and Shane Sims for their help on the tool implementation of this research. We are grateful to Bram Adams for helpful discussions on a former version of the paper. This research was partially supported by the Natural Sciences and Engineering Research Council of Canada, NSERC Discovery Grant 250343-12 and Alberta Innovates Technology Future. Last but not least, we highly appreciate the thoughtful comments and suggestions made by the anonymous reviewers and the handling editor.

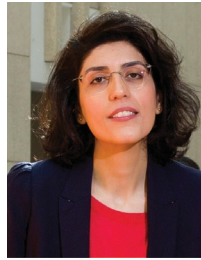

**Maleknaz Nayebi** is an Assistant Professor at Ecole Polyetchnique of Montreal. She received her PhD degree from at the Software Engineering Decision Support lab from The University of Calgary in Canada. The PhD was on *Analytical Release Management for Mobile Apps*. She has six years of professional software engineering experience. Her main research interests are in mining software repositories, release engineering, open innovation and empirical software engineering. Maleknaz co-chaired RE data track 2018, IWSPM 2018, IASESE 2018 advanced school, and OISE 2015. Maleknaz is a member of the IEEE and ACM. Her homepage is http://maleknazn.com.

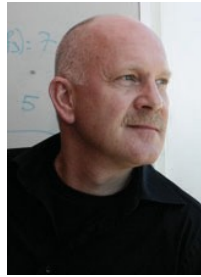

**Guenther Ruhe** is the Industrial Research Chair in Software Engineering at the University of Calgary. His research focuses on product release planning, software project management, decision support, data analytics, empirical software engineering, and search-based software engineering. He is the editor in chief of Information and Software Technology and was the General Chair of the Requirements Engineering conference RE18. He is a senior member of IEEE and a member of ACM.

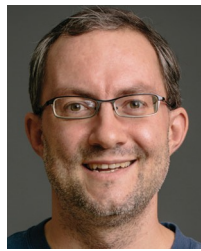

**Thomas Zimmermann** is a Senior Researcher at Microsoft. He received his Ph.D. degree from the Saarland University in Germany. His research interests include empirical software engineering, data science, and mining software repositories. His work focuses on software productivity, software analytics, and recommender systems for software development. He is co-editor in chief of the Empirical Software Engineering journal. He is a senior member of the IEEE. His homepage is http://thomas-zimmermann.com.

# APPENDIX I: SAMPLE SYNTHESIS

## 1 DATA AND MAIN STEPS OF SYNTHESIS

In this Appendix, for a smaller data set, we illustrate the execution steps of synthesis process. We considered the third application as presented in Section 4 and selected a subset of 466 the original set of 6,003 apps. From applying GWM, these itemsets (sequences of release cycle times) were categorized as below:

- ML including 22 apps,
- (ML)* including 112 apps,
- (M*L) including 154 apps,
- (M*L)* including 56 apps,
- M*L* including 102 apps, and
- (M*L*)* including 281 apps.

The relation between these regular expressions is shown in Figure 1.

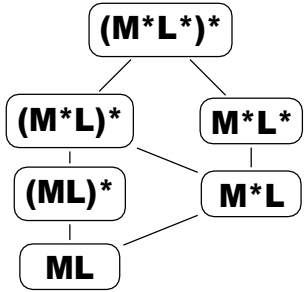

Fig. 1. The hierarchy of four regular expressions representing 941 sample apps.

We showed the main synthesize steps on the treatments in Figure 1 within Table 1. "Significant" in the test result column shows that the p-value of the Mann-Whitney test for the two is less than the stated threshold. As the result of this synthesis process nodes (M*L*)* and M*L remain within the hierarchy.

## 2 DISCUSSION

To elaborate more on the reasoning of the GWM findings, we also provide a sample of (i) a correct (✓) and of (ii) a wrong (✗) inference:

✗ **Wrong interpretation**: One should release her/his app within consecutive short cycles ($s*$) to receive better rating. GWM found that $S*$ affects app rating significantly and higher than other release sequences, but this does not mean causality. GWM works in the observation space rather than in the theory space. We cant state that these apps are getting higher ratings because they follow consecutive short cycles. Short release cycles might be one of the reasons. They should be considered in the models to predict apps success.

✓ **Correct interpretation**: A repetitive sequence of short release cycles followed by a long cycle $(SL)*$ does not impact app rating significantly different in comparison to a strategy such as $S*L*$ or $(S*L*)*$. The app developer is following $(SL)*$ to release her alpha and beta versions of her app. She released an alpha version and she planned to release the beta within the next six days (short cycle). After two days, she receives several negative comments on app defects. She cant fix all the bugs in the remaining days and is thinking about changing the schedule and release the app later. Using the GWM results by analyzing former experience, she does not need to be concerned about the negative affect of deadline extension on customer satisfaction.

TABLE 1: Synthesis process steps for Figure 1.

| Step | Node | Selection criteria | Parent(s) | P-value | Test results | Action |
|---|---|---|---|---|---|---|
| 1 | ML | Node with distance 3 and found by DFS | (ML)*, M*L | 0.025 | (ML)*, M*L := insignificant | Merge ML with (ML)* |
| 2 | (ML)* | Node with distance 2 and found by DFS | (M*L)* | 0.017 | (ML)*, (M*L)* := insignificant | check its siblings. |
| 3 | M*L | Sibling of an insignificant node | (M*L*)* | 0.012 | M*L, (M*L*)* := significant | compare with siblings and merge (ML)* with its parent (M*L)* |
| 4 | (M*L)* | Node with distance 1 and found by DFS | (M*L*)* | 0.008 | (M*L)*, M*L =: significant (M*L)*, (M*L*)* =:insignificant | Check its siblings |
| 5 | M*L* | Sibling of an insignificant node | (M*L*)* | 0.006 | M*L*, M*L =: significant (M*L)*, (M*L*)* =:insignificant | Merge (M*L)* with its parent (M*L*)* because of smaller effect size and bigger p-value |
| 6 | M*L* | Undecided node in distance 1 | (M*L*)* | 0.005 | M*L*, (M*L*)* =: insignificant | Merge M*L* with its parent (M*L*)* and transfer M*L to (M*L*)* |
| 7 | M*L | Node in distance 1 | (M*L*)* | 0.004 | M*L, (M*L*)* =: significant | Terminate synthesis process. |